\documentclass[11pt,a4paper]{article}
\pdfoutput=1
\usepackage[utf8]{inputenc}

\usepackage{jheppub}
\usepackage{graphicx,amssymb,amsmath,amsfonts}
\usepackage{amsthm}
\usepackage[english]{babel}
\usepackage{hyperref}
\usepackage{cleveref}
\usepackage{subcaption}

\usepackage{braket}

\newcommand{\mn}{{\mu\nu}}
\newcommand{\rs}{{\rho\sigma}}
\newcommand{\mnrs}{{\mu\nu\rho\sigma}}

\newcommand{\V}{\mathbf}
\newcommand{\rmd}{{\mathrm{d}}}
\newcommand{\p}{{\partial}}
\newcommand{\vep}{\varepsilon}

\newcommand{\zb}{{\bar{z}}}
\newcommand{\wb}{{\bar{w}}}
\newcommand{\gzz}{{\gamma_{z\bar{z}}}}

\newcommand{\gww}{{\gamma_{w\bar{w}}}}

\newcommand{\mA}{\mathcal{A}}

\newcommand{\mL}{\mathcal{L}}

\newcommand{\td}[1]{\widetilde{d^3 #1}\,}

\newcommand{\hA}{{\hat{A}}}

\newcommand{\w}{{\omega}}

\newcommand{\hc}{\text{h.c.}}
\newcommand{\tin}{{\text{in}}}

\newcommand{\Di}{{\frac{1}{\Delta}}}
\newcommand{\Dis}{{\frac{1}{\Delta_s}}}

\makeatother

\title{Soft Photon Hair on Schwarzschild Horizon from a Wilson Line Perspective}
\author{Sangmin Choi and Ratindranath Akhoury}

\affiliation{Leinweber Center for Theoretical Physics, \\
Randall Laboratory of Physics, Department of Physics,\\
University of Michigan, Ann Arbor, MI 48109, USA}

\emailAdd{sangminc@umich.edu, akhoury@umich.edu}

\abstract{
We develop a unified framework for the construction of soft dressings at boundaries of spacetime, such as the null infinity of Minkowski spacetime and the horizon of a Schwarzschild black hole. The construction is based on an old proposal of Mandelstam for quantizing QED and considers matter fields dressed by Wilson lines. Along time-like paths, the Wilson lines puncturing the boundary are the analogs of flat space Faddeev-Kulish dressings. We focus on the Schwarzschild black hole where our framework provides a quantum-field-theoretical perspective of the Hawking-Perry-Strominger viewpoint that black holes carry soft hair, through a study of the Wilson line dressings, localized on the horizon.
}

\begin{document}
{
\maketitle
}

\section{Introduction}\label{sec:introduction}

Hawking's discovery of black hole evaporation \cite{Hawking:1974sw} more than forty years ago brought with it the information loss problem \cite{Hawking:1976ra} in the context of the semiclassical theory. A perhaps related problem is that of developing an understanding of black hole entropy in terms of microscopic degrees of freedom. The recent proposal of Hawking, Perry and Strominger \cite{Hawking:2016msc} has provided hope that it contains elements that could provide for a better understanding of both of these open problems in black hole physics. This work itself was motivated by earlier papers of Strominger and collaborators \cite{Strominger:2013jfa,He:2014laa,He:2014cra,Strominger:2014pwa,Kapec:2015vwa,Kapec:2015ena} (see \cite{Strominger:2017zoo} for a review) where, in the context of flat Minkowski space, it was shown that, the three apparently disparate subjects of asymptotic symmetries, soft theorems and memory effect are in fact different aspects of the same physical description. The contribution of the Hawking, Perry and Strominger paper \cite{Hawking:2016msc} was to introduce new elements that were not considered in the original analysis of information loss in \cite{Hawking:1976ra}. In particular, there are two novel ingredients in the proposal of \cite{Hawking:2016msc}. First, they discovered asymptotic symmetry groups which act simultaneously on the future and past boundaries of spacetime containing a black hole. The corresponding conserved charges associated with these asymptotic symmetries can then constrain the Hawking radiation in a non-trivial manner. Thus, one is able to distinguish black holes by these charges in addition to their mass, electric charge and angular momentum -- in violation of the classical no hair theorems. Secondly, the infalling matter particles passing through the horizon induce there the asymptotic symmetry transformation thereby creating the soft hair. Thus, there exist a large degeneracy of black holes having the same mass but different numbers of soft hair on the horizon.  Whether these new ingredients are sufficient to resolve completely the information loss problem or the problem of microstates is an open question which will not be considered further in this paper. In fact, our focus here will be to study the conclusions of \cite{Hawking:2016msc}, for the case of electromagnetism, from a different perspective, namely, we will deduce the existence and study the properties of the soft hair on the horizon of a black hole within the framework of a quantum mechanical, gauge-invariant but path-dependent formulation of QED proposed by Mandelstam \cite{Mandelstam:1962mi} where Wilson line dressed matter fields were first introduced.\footnote{
	Dirac also approaches QED from a similar perspective, see \cite{Dirac}.}
Through our work, we hope to convince the reader that the Wilson line is an effective tool to study the soft hair at
	both infinity and the horizons.

 In Mandelstam's formulation, a (spacetime) point is a derived concept, regarded as the end of a path. The theory is then formulated in the space of all paths. In order to implement this, Mandelstam introduces what we would now call a ``Wilson line". Conventional matter fields at, say, a spacetime point $x$ are replaced by ones that are dressed by Wilson lines extending from the past time-like infinity to the point $x$ along a path. This formulation of QED is gauge-invariant but path-dependent and does not involve potentials in any other way. The reader is referred to \cite{Mandelstam:1962mi} for details and comparison with the usual formulation. What is of relevance for us in this paper are the dressed matter fields which will play a fundamental role in that they provide a natural framework to study the generation and properties of the soft ``hair" at the spacetime boundaries. For the purposes of studying infrared dynamics in flat space, a great simplification occurs in that only a particular path in the Wilson line becomes relevant. Specifically, what is of interest are the Wilson lines for which the path is time-like and linear in proper time. This is because we are interested in the asymptotic dynamics at time-like infinity (for massive particles) or at null infinity (for massless particles) where particle paths are straight lines in the leading approximation. We will distinguish when the point $x$ is anywhere in the bulk of spacetime, and when it is at the spacetime boundaries. When it is at the spacetime boundary, it will be referred to as a Wilson line puncture. We will see in the subsequent sections that the edge or boundary gauge modes relevant for the Wilson line (the dressing) will be responsible for generating the soft photon ``hair". A natural question that arises is how is this Wilson line dressing for asymptotic particles related to the Faddeev-Kulish dressing introduced in \cite{Kulish:1970ut}  in order to obtain infrared finite S matrix elements. It was noted in \cite{Jakob:1990zi,Zwanziger:1973mp} that in flat Minkowski space, for the straight line time-like path, the Wilson line dressing and the Faddeev-Kulish one are the same. In this paper, we first further quantify this connection and show the consistency of this perspective with that of \cite{Strominger:2017zoo, Kapec:2017tkm}. The Wilson line dressing generates a large gauge transformation at time-like infinity and the associated charge satisfies the usual canonical commutation relations. Additionally, the Wilson line dressing acting on the vacuum state gives another vacuum with a different value of the large gauge transformation charge. In this way one has a large degeneracy of vacua containing different numbers of soft photons.
 
 This same perspective is next applied to curved spacetimes. First we consider the right Rindler wedge and then the Schwarzschild spacetimes where now in addition to the boundaries at infinity we also have the past and future horizons. The new feature is the structure of the edge modes and of the Wilson line dressings at the horizon. For the right Rindler wedge, we focus on the Wilson line for a time like path which punctures the future horizon.  Our analysis again identifies these punctures at the horizon with the analogue of the Faddeev-Kulish dressing and show that it carries a definite large gauge transformation charge at the horizon. This charge is explicitly constructed and has the appropriate canonical commutation relations with the Wilson line. Since the horizon is an infinite red-shift surface, the charge, as expected, is static. Once again there is a large degeneracy of the vacuum states representing states with different numbers of zero-energy photons. The case of the right Rindler wedge has been recently considered from a different viewpoint in \cite{Blommaert:2018rsf,Blommaert:2018oue} where the edge Hilbert space was constructed in the Lorenz gauge. Using the canonical quantization procedure developed in \cite{Lenz:2008vw}, the Wilson line punctures on the black hole horizon, as well as the large gauge transformation charges are constructed next and the interrelationship between them is established. The Wilson line punctures can be used to construct the edge Hilbert space. The bulk state is one of a black hole with a mass $M$ which is degenerate requiring the soft large gauge transformation charge encoded in the edge Hilbert space to correctly identify the complete state. The edge and the bulk Hilbert space are factorized, however, because of this degeneracy the soft hairs have observable consequences, manifest for example in the memory effect \cite{Bieri:2013hqa,Strominger:2014pwa}. It is also worth noting here that static nature of the charges in the case of Rindler and Schwarzschild indicate a difference with the flat space Minkowski results. This circumstance allows for a state counting criteria to be applied to the charges on the horizon as distinct from the case of the Wilson line punctures on other spacetime boundaries \cite{Blommaert:2018oue}.

 This paper is organized as follows: Section \ref{sec:minkowski} discusses the flat space case and Section \ref{sec:rindler} is devoted to the analysis of the soft photon hair on the right Rindler wedge. Section \ref{sec:schwarzschild} deals directly with the soft hair on the horizon of Schwarzschild spacetimes and section \ref{sec:discussion} contains a discussion of our results. Certain technical details related to section \ref{sec:schwarzschild} are relegated to an appendix.

\section{Wilson lines and soft charge in Minkowski spacetime}\label{sec:minkowski}

This section is aimed at reviewing topics in flat Minkowski spacetime that will be crucial in our subsequent analysis of Rindler and Schwarzschild horizons.
In the first subsection, selected materials from \cite{Mandelstam:1962mi} and \cite{Jakob:1990zi} are used
	to show that the Faddeev-Kulish dressings of asymptotic states \cite{Kulish:1970ut,Ware:2013zja}
	are in fact Wilson lines along a specific time-like path.
Then in the second subsection, we explore the connections between Faddeev-Kulish dressings, Wilson line punctures, edge states and
 surface charges associated with the asymptotic symmetry transformation developed in \cite{Kapec:2017tkm}.

\subsection{Equivalence of Wilson lines and Faddeev-Kulish dressings}\label{sec:mink1}

A gauge-invariant formulation of QED using path-dependent variables dates back to Mandelstam's work \cite{Mandelstam:1962mi}.
In this formulation, the conventional matter fields are dressed by Wilson lines extending out to infinity.
It is known \cite{Zwanziger:1973mp, Jakob:1990zi} that taking the path in each Wilson line
	to be the time-like path of an asymptotic particle yields the Faddeev-Kulish dressings \cite{Kulish:1970ut}.
In this section we briefly review this connection.

Under a gauge transformation, the gauge and matter fields transform as
\begin{align}
	\varphi(x) &\to e^{-ie\Lambda(x)}\varphi(x),
	\\
	A_\mu(x) &\to A_\mu(x) + \p_\mu\Lambda(x).
\end{align}
Mandelstam introduces a non-local path dependent variable which is the matter field dressed with a Wilson line, i.e., 
\begin{align}\label{psi}
	\Psi(x\,|\,\Gamma) = \mathbb{P}\exp\left\{ie\int^x_\Gamma d\xi^\mu A_\mu(\xi)\right\}\varphi(x),
\end{align}
along the path $\Gamma$, with the path-ordering operator $\mathbb{P}$. 
We will show that this path-dependent dressing describes the Faddeev-Kulish dressing for a particular time-like, straight line path relevant for 
asymptotic field.
In the Lorenz gauge, the equation of motion of the gauge field with source $J_\mu$ is
\begin{align}
	\square A_\mu(x) = J_\mu(x).
\end{align}
Therefore one may use the retarded Green's function $G_\text{ret}$ to decompose
\begin{align}\label{Yang}
	A_\mu(x) &= A^\tin_\mu(x) + \int d^4z\, G_\text{ret}(x-z)J_\mu(z),
\end{align}
where the $G_\text{ret}$ solves $\square G_\text{ret}(x) = \delta^{(4)}(x)$,
	and the incoming asymptotic field $A_\mu^\tin$ is the homogeneous solution satisfying
	$ \square A^\tin_\mu(x) = 0$.
Then, the dressing in \eqref{psi} can be written as
\begin{align}
	\mathbb{P}\exp\left\{ie\int^x_\Gamma d\xi^\mu A_\mu(\xi)\right\}
	&= \mathbb{T}\exp\left\{ie\int^x_\Gamma d\xi^\mu A^\tin_\mu(\xi)
		+ ie\int^x_\Gamma d\xi^\mu \int d^4z\, G_\text{ret}(\xi-z)J_\mu(z)\right\}
	\\ &= \exp\left\{ie\int^x_\Gamma d\xi^\mu A^\tin_\mu(\xi)\right\} \times (\text{phase factors}),
\end{align}
where $\mathbb{P}$ is replaced by the time-ordering operator $\mathbb{T}$ since $\Gamma$ is time-like.
The path-ordering has been removed at the price of gaining an infinite c-number phase,
	which is related to the Coulomb phase and thus is not of interest for this paper.
We anticipate letting $x^0\to-\infty$ for an asymptotic incoming particle.
With this limit in mind, we may assume $\Gamma$ to be the trajectory of a free particle described by a constant four-velocity $v^\mu$,
	and parametrize $\xi^\mu = x^\mu + \tau v^\mu$.
Then,
\begin{align}
	\exp\left\{ie\int^x_\Gamma d\xi^\mu A^\tin_\mu(\xi)\right\}
	&= \exp\left\{ie\int^0_{-\infty} d\tau \frac{d\xi^\mu}{d\tau} A^\tin_\mu(\xi)\right\}
	\\ &= \exp\left\{ie v^\mu\int^0_{-\infty} d\tau A^\tin_\mu(x+v\tau)\right\}.
\end{align}
For the in field, we have the standard asymptotic mode expansion
\begin{align}
	A_\mu^\tin(x) = \int \td{k}
		\left[
			a_\mu(\V k) e^{ik\cdot x} + a_\mu^\dagger(\V k) e^{-ik\cdot x}
		\right],
\end{align}
where $\td{k}=\frac{d^3k}{(2\pi)^3(2\w)}$ with $\w = |\V k|$ is the Lorentz-invariant measure, and
\begin{align}
	a_\mu(\V{k}) = \sum_{\ell=\pm} \epsilon^{\ell*}_\mu(\V{k})a_\ell(\V{k}),
	\quad a_\mu^\dagger(\V{k}) = \sum_{\ell=\pm} \epsilon^{\ell}_\mu(\V{k})a^\dagger_\ell(\V{k}),
\end{align}
with the polarization tensor $\epsilon_\mu^\ell(\V k)$.
The creation and annihilation operators satisfy the standard commutation relations
\begin{align}
	\left[a_{\ell}(\V k), a_{\ell'}^\dagger(\V k')\right] = \delta_{\ell\ell'} (2\pi)^3 (2\w) \delta^{(3)}(\V k - \V k').
	\label{firstCR}
\end{align}
Using the mode expansion, we may write
\begin{align}
	v^\mu\int^0_{-\infty} d\tau A^\tin_\mu(x+v\tau)
	&= v^\mu\int^0_{-\infty} d\tau
		\int \td{k}
		\left[
			a_\mu(\V k) e^{ik\cdot (x + v\tau)} + \hc
		\right]
	\\ &= -i\int \td{k} \frac{p^\mu}{p\cdot k}
		\left[
			a_\mu(\V k) e^{ik\cdot x} - \hc
		\right],
\end{align}
where $p^{\mu}=mv^{\mu}$, and we have used the boundary condition \cite{Kulish:1970ut}
\begin{align}\label{boundarycondition}
	\int^0_{-\infty} d\tau \,e^{ik\cdot v\tau} = \frac{1}{ik\cdot v}.
\end{align}
Therefore, we obtain
\begin{align}
	\exp\left\{ie\int^x_\Gamma d\xi^\mu A^\tin_\mu(\xi)\right\}
	&=\exp\left\{-e
		\int \td{k}\frac{p^\mu}{p\cdot k}
		\left (a_\mu^\dagger(\V{k}) e^{-ik\cdot x} - a_\mu(\V{k}) e^{ik\cdot x}\right)
	\right\}.
\end{align}
Now, recall that we anticipate $x^0\to-\infty$.
Under this limit, non-vanishing contribution to the integral comes only from $k\to 0$ by virtue of the Riemann-Lebesgue lemma.
Following the construction of \cite{Kulish:1970ut}, we implement this by replacing $e^{\pm ik\cdot x}$ with
	a scalar function $\phi(p,k)$ having support in a small neighborhood of $k=0$ and satisfying
	$\phi\to 1$ as $k\to 0$.
Then, we may write,
\begin{align}
	\exp\left\{ie\int^x_\Gamma d\xi^\mu A^\tin_\mu(\xi)\right\}
	&=W(\V p),
	\label{wleqfk}
\end{align}
where $W(\V p)$ is, up to a unitary transformation, the Faddeev-Kulish operator (or dressing) of an asymptotic incoming particle of momentum $\V p$,
\begin{align}
	W(\V p) &=\exp\left\{-e
		\int \td{k} \frac{p^\mu}{p\cdot k}\phi(p,k)
		\left (a_\mu^\dagger(\V{k}) - a_\mu(\V{k})\right)
	\right\}.
\end{align}
Equation \eqref{wleqfk} shows that a Wilson line along the trajectory of an asymptotic particle
	corresponds to the Faddeev-Kulish dressing.

\subsection{Faddeev-Kulish dressings and the soft charge}

\begin{figure}
	\centering
	\includegraphics[width=.45\textwidth]{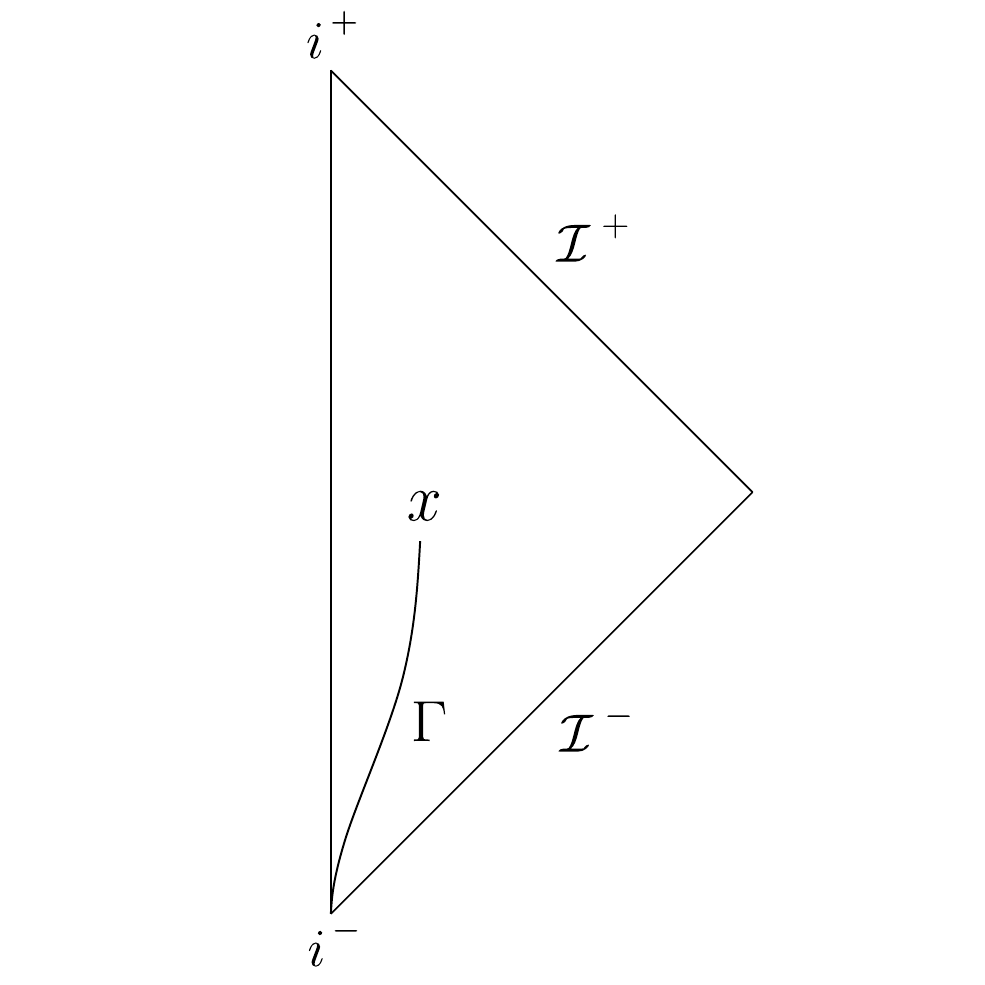}
    \caption{
    	A Penrose diagram of the Minkowski spacetime, where $\mathcal{I}^+$ ($\mathcal{I}^-$) represents the future (past) null infinity
    		and $i^+$ ($i^-$) represents the future (past) time-like infinity.
    	The spacetime point $x$ is the position of a massive dressed particle,
    		and the time-like path $\Gamma$ extends from $i^-$ to $x$.
    	The time-like infinities $i^\pm$ are 3-dimensional hyperbolic spaces $\mathbb{H}_3$
    		each parametrized by a 3-vector, see for example \cite{Strominger:2017zoo}.
    }
    \label{fig:mink}
\end{figure}

In the previous subsection, we have seen that Faddeev-Kulish dressings are essentially Wilson lines.
Let us consider a Wilson line along a time-like curve $\Gamma$ of constant momentum $p$ ending at a point $x$,
	as in figure \ref{fig:mink}.
The Wilson line stretches all the way down to the past time-like infinity $i^-$, where the asymptotic phase space is the hyperbolic space $\mathbb{H}_3$.
If this Wilson line is dressing an asymptotic massive charged particle, as is the case under our consideration, we can assume that
	the limit $x^0\to -\infty$ is being taken.
In this picture, one can see that the Faddeev-Kulish dressing
	can essentially be viewed as a Wilson line puncture on $i^-$, the asymptotic boundary of Minkowski spacetime.
The term ``puncture" will henceforth be used to denote a Wilson line along a time-like path piercing the spacetime
	boundary of our interest.

Let us put this intuitive description on a more formal ground.
The Minkowski spacetime has the following metric in terms of the Cartesian coordinates
\begin{align}
	ds^2 &= -dt^2 + dx_1^2 + dx_2^2 + dx_3^2.	
\end{align}
We introduce the advanced set of coordinates $(v,r,z,\zb)$, which is related to the Cartesian coordinates by
\begin{align}
	v = t + r,\qquad
	r^2 = x_1^2 + x_2^2 + x_3^2,\qquad
	z = \frac{x_1+ix_2}{r+x_3}.
\end{align}
The Minkowski metric can then be written as
\begin{align}
	ds^2 &=-dv^2 + 2dvdr + 2r^2 \gzz dzd\zb,
\end{align}
where $\gzz = 2/(1+z\zb)^2$ is the unit 2-sphere metric.
In terms of these coordinates, the momentum measure is $d^3k = \w^2 d\w \gzz d^2z$, so we may write the asymptotic gauge field as
\begin{align}
	A^\tin_\mu(x)
		&= \int \frac{d^3k}{(2\pi)^3}\frac{1}{2\w}\left [a_\mu(\V{k}) e^{ik\cdot x}+a_\mu^\dagger(\V{k}) e^{-ik\cdot x}\right ]
		\\ &= \frac{1}{16\pi^3}\int \w d\w\gzz d^2z\left [a_\mu(\w \hat{\V k}) e^{ik\cdot x}+a_\mu^\dagger(\w \hat{\V k}) e^{-ik\cdot x}\right ].
\end{align}
Here $\hat {\V k}$ is a unit 3-vector that points in the direction defined by $(z,\zb)$, such that $\V k = \w \hat{\V k}$.
Let us employ the usual polarization tensors
\begin{align}
	\epsilon^{+\mu} = \frac{1}{\sqrt{2}}(\zb,1,-i,-\zb),
	\quad
	\epsilon^{-\mu} = \frac{1}{\sqrt{2}}(z,1,i,-z),
\end{align}
the plane-wave expansion,
\begin{align}
	e^{ik\cdot x}
		&= 4\pi e^{-i\omega t}\sum_{\ell=0}^\infty i^{\ell} j_{\ell}(\omega r)\sum_{m=-\ell}^\ell Y_{\ell m}(\hat{\V{k}})Y_{\ell m}^*(\hat{\V{x}})
		\\ &= \frac{2\pi i}{\omega r}e^{-i\omega v}\delta^{(2)}(\hat{\V{k}} + \hat{\V{x}}) + O\left(r^{-2}\right),
\end{align}
as well as the relation $A_z = \p_z(x^\mu A_\mu)$ to obtain the non-vanishing components \cite{Kapec:2017tkm}:
\begin{align}
	A_z(v,z,\zb) &= \lim_{r\to \infty} A_z(v,r,z,\zb)
	\\ &= \frac{i}{8\pi^2}\sqrt{\gzz}\int d\w\left (a_-^\dagger(-\w \hat{\V{x}}) e^{i\w v} - a_+(-\w\hat{\V x})e^{-i\w v}\right ).
	\label{Az}
\end{align}
Here the expression $a_\pm(-\w \hat{\V x})$ should be understood as a particle operator with $\w>0$ with momentum in the direction $-\hat{\V x}$.
The minus sign comes from the fact that a massless particle moving in the direction $(z,\zb)$ will be mapped to its antipodal point
	in the past infinity.

Observe in \eqref{Az} that taking the limit $v\to -\infty$ forces the integral to get contributions only from the zero-modes $\w=0$
	by virtue of the Riemann-Lebesgue lemma.
We may use the method of \cite{Kulish:1970ut,Ware:2013zja} to implement this explicitly, by introducing an infrared
	scalar function $\phi(\w)$ that has support only in a small neighborhood of $\w=0$ and satisfies $\phi(0)=1$:
\begin{align}
	A_z(z,\zb) = \frac{i}{8\pi^2}\sqrt{\gzz}\int dw \left (a_-^\dagger(-\w\hat{\V{x}}) - a_+(-\w\hat{\V{x}})\right )\phi(\w).
	\label{Azphi}
\end{align}
A Faddeev-Kulish dressing $W(\V p)$ of a particle with momentum $\V p$ can be written in terms of these boundary modes.
\begin{align}
	W(\V p)
	&= \exp\left\{\frac{ie}{2\pi}
		\int d^2z\sqrt{\gzz}\left  [
		\left(
			\frac{p\cdot \epsilon^-}{p\cdot \hat k}
		\right)A_z(z,\zb)
		+\left(
			\frac{p\cdot \epsilon^+}{p\cdot \hat k}
		\right)A_\zb(z,\zb)
		\right  ]
	\right \},
	\label{W1}
\end{align}
where $\hat{k}^\mu = (1,\hat{\V k})=(1,-\hat{\V x})$.

In terms of the language used in \cite{Blommaert:2018rsf}, the edge modes $A_z(z,\zb)$ and $A_\zb(z,\zb)$ are the
	zero modes that exponentiate to the Wilson line sourced and localized at the boundary. In this reference which deals with the case 
of Rindler space, for each edge mode annihilation operator $a_\V{k}$, there is a conjugate variable $q_\V{k}$ such that
\begin{align}
	[a_\V{k}, q_{-\V{k}'}] = i\delta_{\V k\V{k}'},
\end{align}
The eigenspace of this conjugate variable is more natural (compared to the eigenspace of $a_\V{k}$) 
in the sense that it diagonalizes the boundary Hamiltonian of the Rindler space. 
We will see below that even in flat Minkowski space, the vacua of Faddeev-Kulish states define an eigenspace analogous to that of $q_\V{k}$ in a manner which was
previously discussed in \cite{He:2014cra}.

For this purpose, consider a function $\vep(z,\zb)$ on the 2-sphere.
The conserved charge $Q_\vep$ associated with this function can be written as \cite{He:2014cra}
\begin{align}
	Q_\vep = Q_\vep^\text{soft} + Q_\vep^\text{hard},
\end{align}
where the hard charge $Q_\vep^\text{hard}$ contains charged matter current and hence commutes with
	the boundary field $A_z(z,\zb)$, and the soft charge $Q_\vep^\text{soft}$ is given by
\begin{align}
	Q^\text{soft}_\vep = -2\int d^2z \,\p_\zb\vep(z,\zb)N_z(z,\zb) = -2\int d^2z\, \p_z\vep(z,\zb)N_\zb(z,\zb)
	\label{Qs_Nz}.
\end{align}
Here the operator $N_z(z,\zb)$ is defined as
\begin{align}
	N_z(z,\zb) = \int_{-\infty}^\infty dv\, \p_v A_z(z,\zb),
	\label{Nz}
\end{align}
and contains only zero-energy photon operators, as one can see from the expression
\begin{align}
	N_z(z,\zb)
		&= - \frac{1}{4\pi}\sqrt{\gzz} 
			\int_0^\infty d\w\,\w\,\delta(\w)\left ( a_-^\dagger(-\w \hat{\V{x}}) + a_+(-\w \hat{\V{x}})\right ),
	\label{Nzexpansion}
\end{align}
obtained by using the mode expansion \eqref{Azphi} and the integral representation
\begin{align}
	\label{intrepdelta}
	\int_{-\infty}^\infty d\w\,e^{\pm i\w v} = 2\pi\delta(\w).
\end{align}
From \eqref{Azphi} and \eqref{Nzexpansion}, we obtain by direct calculation
\begin{align}
	[A_z(z,\zb),N_\wb(w,\wb)]
		&= \frac{i}{2}\sqrt{\gzz\gww}
		\int d\w\, d\w'\,\w\phi(\w)\w'\delta(\w')
		\delta^{(3)}(\w\hat{\V{x}}_z - \w'\hat{\V{x}}_w),
\end{align}
where we used the commutation relation \eqref{firstCR}.
Now, note that we may write
\begin{align}
	\delta^{(3)}(\w\hat{\V{x}}_z - \w'\hat{\V{x}}_w) = \frac{1}{\w^2 \gzz}\delta(\w-\w')\delta^{(2)}(z-w).
\end{align}
Therefore, with the convention
\begin{align}
	\int_0^\infty \,d\w\,\delta(\w)f(\w) = \frac{1}{2}f(0),
\end{align}
of delta functions acting on the boundary of the integration domain, we obtain the commutation relation
\begin{align}
	[A_z(z,\zb),N_\wb(w,\wb)]
		&= \frac{i}{2}\delta^{(2)}(z-w).
		\label{comm}
\end{align}
It follows from the expression \eqref{Qs_Nz} that
\begin{align}
	\left[
		Q_\vep, A_z(z,\zb)
	\right]
	&= i\p_z\vep(z,\zb).
	\label{W2}
\end{align}
It is worth noting here that $Q_\vep$ may be replaced by $Q_\vep^\text{soft}$ since $Q_\vep^\text{hard}$ commutes with $A_z(z,\zb)$.
There is an immediate consequence of \eqref{W2} and \eqref{W1}, which has been pointed out in \cite{Kapec:2017tkm}.
Consider a vacuum $\ket{0}$ such that $Q_\vep\ket{0}=0$
	and dress it with the Faddeev-Kulish operator to construct a state $W(\V p)\ket{0}$.
The expression \eqref{W1} shows that $W(\V p)$ involves only the boundary gauge fields and thus only the zero-mode photon operators,
	qualifying $W(\V p)\ket{0}$ as a vacuum.
The two vacua $\ket{0}$ and $W(\V p)\ket{0}$ are distinct, since the latter carries soft charge,
\begin{align}
	Q_\vep W(\V p)\ket{0}
	&= -\frac{e}{2\pi}\int d^2z\sqrt{\gzz}
		\left\{
			\left(\frac{p\cdot \epsilon^-}{p\cdot \hat k}\right)\p_z \vep(z,\zb)
			+ \left(\frac{p\cdot \epsilon^+}{p\cdot \hat k}\right)\p_\zb \vep(z,\zb)
		\right\}
		W(\V p)\ket{0},
\end{align}
as implied by \eqref{W2}.
Since there are infinitely many dressings $W(\V p)$,
	there exists an infinite number of degenerate vacua, each characterized by its soft charge.
The selection rule arising from the charge conservation manifests itself as the infrared divergence of scattering amplitudes,
	and the asymptotic states of Faddeev and Kulish are the eigenstates of the conserved charge $Q_\vep$.
This has been investigated for the flat spacetime both in QED \cite{Kapec:2017tkm}
	and in perturbative gravity \cite{Choi:2017ylo}.
	
In summary, we have seen that the flat-space Faddeev-Kulish dressings can be written as Wilson line punctures
	on the asymptotic boundary of Minkowski spacetime.
The massless gauge field has non-vanishing components at the asymptotic boundary, and the Wilson line can be written
	as a linear combination of these fields.
On the other hand, the soft charge of the large gauge symmetry is a linear combination of a variable which is canonically conjugate to the
	boundary gauge field.
As a consequence, each dressing carries a definite soft charge, parametrized by a 3-momentum $\V p$.
The dressings carry zero energy, and therefore can be used to generate a Hilbert space consisting of an infinite number of distinct vacua.
This space is referred to as the edge Hilbert space in some literature, see for example \cite{Blommaert:2018rsf}.
The soft charge of large gauge transformation is a good quantum number to label the states.
In the next section, we extend this work to the Rindler spacetime and the future Rindler horizon, aiming to draw results consistent
	with the analysis made in \cite{Blommaert:2018rsf}.

\section{Soft hair on future Rindler horizon}\label{sec:rindler}

In the previous section, we have seen that the Faddeev-Kulish dressings of asymptotic states are Wilson lines
	along a time-like path at the future/past time-like infinity.
From this, along with the previous works exploring the connection between large gauge symmetry
	and Faddeev-Kulish dressings \cite{Gabai:2016kuf,Kapec:2017tkm,Choi:2017bna},
	it follows that the set of degenerate vacua carrying soft charge of large gauge transformation
	is obtained by dressing the vacuum with Wilson lines.
Then it is only natural to expect that, in spacetimes exhibiting an event horizon,
	Wilson lines piercing the horizon along a time-like path are the dressings carrying soft hair on the horizon.
In this section, we show that this expectation indeed holds in the Rindler spacetime.
We begin by reviewing a canonical quantization scheme of gauge fields developed in \cite{Lenz:2008vw}.
This will then be used to demonstrate that radial (time-like) Wilson lines in the vicinity of the future horizon
	are the analogue Faddeev-Kulish operators that dress the Fock vacuum to create degenerate vacua carrying soft charge.

Following the notation of \cite{Lenz:2008vw}, we use lower case Latin letters such as $i,j$ to denote the spatial components of tensors
	and capital Latin letters such as $I,J$ to denote the perpendicular components $\V x_\perp$ (that is, $x^2$ and $x^3$).

\subsection{Review of transverse gauge fields in Rindler spacetime}

\begin{figure}
	\centering
	\begin{subfigure}{0.3\textwidth}
	\includegraphics[width=\linewidth]{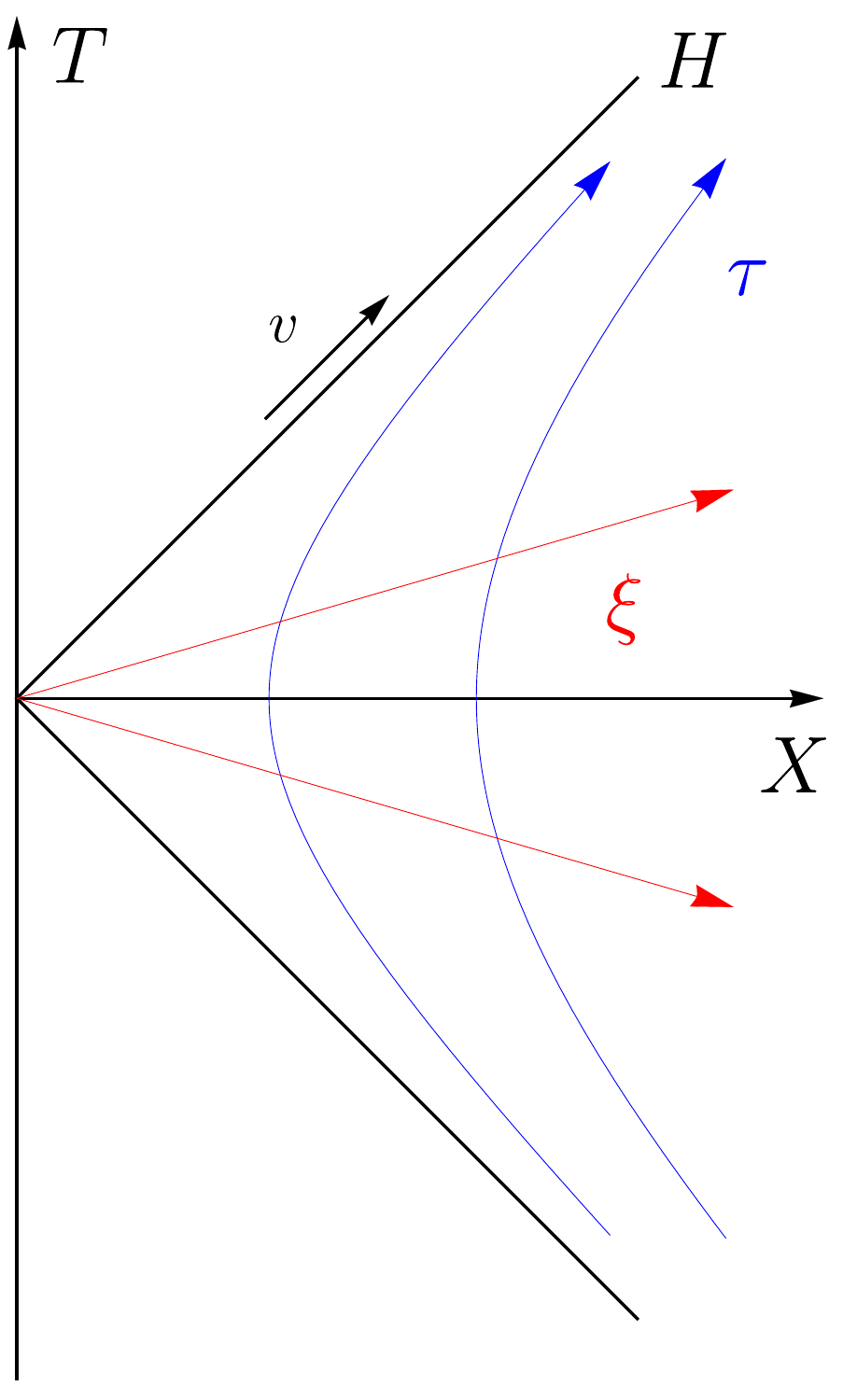}
	\caption{}
	\end{subfigure}
	\hspace{4em}
	\begin{subfigure}{0.3\textwidth}
	\includegraphics[width=\linewidth]{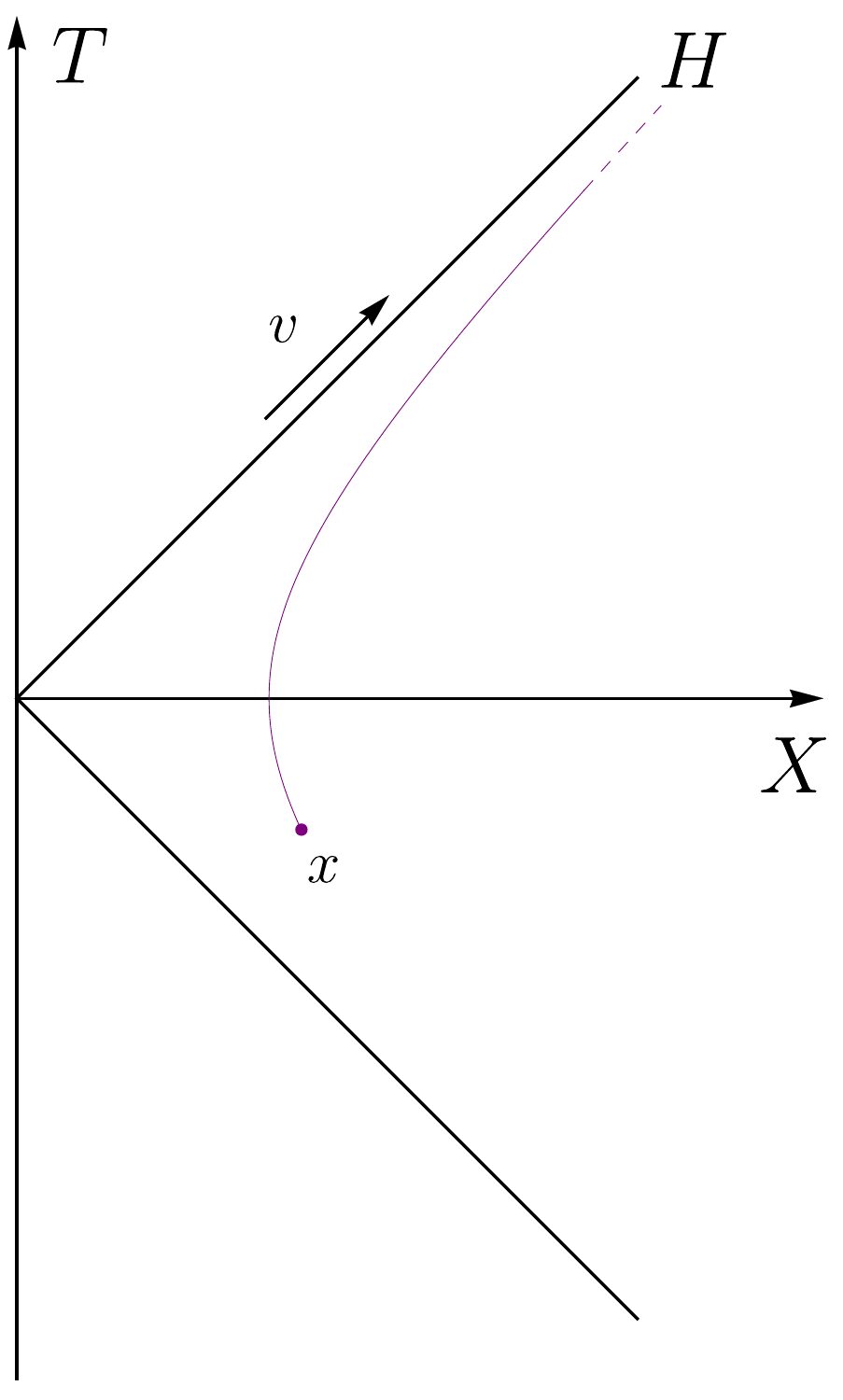}
	\caption{}
	\end{subfigure}
    \caption{
    	A depiction of the Rindler spacetime.
    	The Rindler coordinates ($\tau$, $\xi$) are related to the Minkowski coordinates ($X$, $T$) by
    		$T=\frac{1}{a}e^{a\xi}\sinh(a\tau)$ and
    		$X=\frac{1}{a}e^{a\xi}\cosh(a\tau)$.
    	At the future Rindler horizon $H$, one has $\xi=-\infty$ and $\tau=\infty$.
    	$H$ is parametrized by the advanced time $v=\tau+\xi$, along with the coordinates $\V x_\perp$ which are omitted in the diagram.
    	(a) The constant-$\xi$ curves (marked blue) are parametrized by $\tau$,
    		while the constant-$\tau$ curves (marked red) are parametrized by $\xi$.
    	(b) The purple curve illustrates a Wilson line along a time-like trajectory of a massive particle, starting at a point $x$
    		and extending into $H$.
    }
    \label{fig:rindler}
\end{figure}

Here we will present a brief review the quantization scheme developed in \cite{Lenz:2008vw}.
This quantization in Weyl gauge will especially prove to be useful because it involves only the physical, transverse gauge fields.
The methods introduced here will be relevant to the Schwarzschild case as well.

The Rindler metric takes the following form,
\begin{align}\label{metric}
	ds^2 = e^{2a\xi}(- d\tau^2 + d\xi^2) + d\V{x}_\perp^2.
\end{align}
We will sometimes write $(x^0, x^1, x^2, x^3) = (\tau, \xi, \V x_\perp)$.
The quantization will be carried out in the Weyl gauge,
\begin{align}
	A_0(\tau, \xi, \V x_\perp) = 0.
\end{align}
The Lagrangian density of the gauge field $A_\mu$ coupled to an external current $j^\mu$ is
\begin{align}
	\mL = \sqrt{-g}\left(-\frac{1}{4}F^\mn F_\mn
		- A_\mu j^\mu\right),
\end{align}
where the field strength tensor $F_\mn$ is
\begin{align}
	F_\mn = \p_\mu A_\nu - \p_\nu A_\mu.
\end{align}
The equations of motion are then given by
\begin{align}\label{4eom}
	\p_\mu \sqrt{-g} g^{\mu\rho}g^{\nu\sigma}\left(\p_\rho A_\sigma -\p_\sigma A_\rho\right) = \sqrt{-g} j^\nu,
\end{align}
and the conjugate momenta are
\begin{align}
	\Pi^i = -\sqrt{-g}g^{00}\p_0 A^i.
\end{align}
With the metric of the form \eqref{metric}, the equations of motion reduces to
\begin{align}
	\p_0 \sqrt{-g}g^{00}\p_0 A^i
		+ \p_j \sqrt{-g}g^{jk}g^{il}(\p_k A_l - \p_l A_k) = \sqrt{-g}j^i,
\end{align}
and the Gauss Law ($\nu = 0$ in \eqref{4eom}),
\begin{align}
	\p_i \Pi^i = \sqrt{-g}j^0,
\end{align}
is no longer part of the equations of motion; it becomes a constraint on the conjugate momentum.
Canonical quantization is achieved by postulating the equal-time commutation relation
\begin{align}\label{cano}
	\left[
		\Pi^i(\tau, \xi,\V x_\perp),A_j(\tau,\xi',\V x'_\perp)
	\right]
	&= \frac{1}{i}\delta^i_{j} \delta(\xi-\xi')\delta^{(2)}(\V x-\V x').
\end{align}
Define the transverse projection operator as
\begin{align}\label{proj}
	{P^i}_j = \delta^i_j - \p^i \Di \p_j,
\end{align}
where $\Delta$ is given by
\begin{align}
	\Delta = \p_i \p^i = \p_\xi e^{-2a\xi}\p_\xi + \nabla_\perp^2,
\end{align}
with $\nabla_\perp^2 = \p^2_2 + \p^2_3$.
The projection operator satisfies the following identities,
\begin{align}\label{properties}
	{P^i}_j{P^j}_k = {P^i}_k,\qquad \p_i {P^i}_j = 0,\qquad {P^i}_j \p^j = 0.
\end{align}
Using this, one can define the transverse components of the gauge field and the conjugate momentum which we denote with a hat,
\begin{align}\label{transverse}
	\hA^i = {P^i}_j A^i, \qquad\hat \Pi^i = {P^i}_j \Pi^j.
\end{align}
A nice property of these transverse projections are that by a proper choice of gauge-fixing, the Hamiltonian of the
	gauge field can be formulated completely in terms of the transverse fields \eqref{transverse}, plus a c-number contribution
	describing the effect of the external current
	-- therefore, no unphysical degrees of freedom need to be carried around.
We will not delve into the details of how the dynamics can be written down in terms of the transverse fields;
	we refer the interested readers to \cite{Lenz:2008vw}.

The equal-time commutation relation of the transverse fields can be obtained by transverse projection of the canonical
	relation \eqref{cano}.
To this end, it is convenient to first consider a massless scalar field $\varphi$ in Rindler spacetime, because, as we will later see, the radial gauge fields satisfy the same equation of motion.
The free-field equation of motion is
\begin{align}
	\left(\p_\tau^2 - \Delta_s\right)\varphi = 0,
\end{align}
where the scalar Laplacian $\Delta_s$ is given by
\begin{align}
	\Delta_s = \p_\xi^2 + e^{2a\xi}\nabla_\perp^2.
\end{align}
This can be solved with the ansatz
\begin{align}
	\varphi = e^{-i\w\tau} e^{i\V k_\perp\cdot \V x_\perp} k_{i\frac{w}{a}}(z),
\end{align}
where we defined
\begin{align}
	z = \frac{k_\perp}{a}e^{a\xi},
\end{align}
with $k_\perp = |\V k_\perp|$.
Here $k_{i\frac{w}{a}}$ is the appropriately normalized MacDonald function,
\begin{gather}
	k_{i\frac{\w}{a}}(z) = \frac{1}{\pi}\sqrt{\frac{2\w}{a}\sinh\frac{\pi\w}{a}} K_{i\frac{\w}{a}}(z),
\end{gather}
which forms a complete orthonormal set.
In particular, it satisfies the completeness relation
\begin{align}
	\int_0^\infty d\w\, k_{i\frac{\w}{a}}(z)k_{i\frac{\w}{a}}(z') = az\delta(z-z') = \delta(\xi-\xi').
\end{align}
Among the components of the transverse projection operator \eqref{proj}, the one that will be relevant for our purposes is $i=j=1$.
Using the completeness relations
	and noting that ${P^1}_1$ can be written in terms of the scalar Laplacian $\Delta_s$ as\footnote{
	This can be seen by showing $\Delta_s {P^1}_1 = e^{2a\xi}\nabla_\perp^2$.}
\begin{align}\label{relation}
	{P^1}_1
	&= \Dis e^{2a\xi}\nabla_\perp^2,
\end{align}
one obtains the commutation relation
\begin{align}
	\left[
		\hat \Pi^1(\tau, \xi, \V x_\perp),
		\hA_1(\tau, \xi', \V x'_\perp)
	\right]
	&= -i g_{11} {P^1}_1 g^{11} \delta(\xi-\xi')\delta^{(2)}(\V x_\perp - \V x'_\perp)
	\\ &= -i\int\frac{d^2k_\perp}{(2\pi)^2}z^2 e^{i\V k_\perp\cdot (\V x_\perp - \V x'_\perp)}
		\int_0^\infty d\w \frac{a^2}{\w^2}k_{i\frac{\w}{a}}(z)k_{i\frac{\w}{a}}(z').
		\label{transcomm}
\end{align}
Now, using the properties \eqref{properties} of the transverse projection operator,
	one can show that the equation of motion of the transverse field $\hA_\mu$ with no external current is
\begin{align}
	\sqrt{-g}g^{00}\p^2_0 \hA^i + \p_j \sqrt{-g}g^{jk}g^{il}(\p_k \hA_l - \p_l \hA_k) = 0,
\end{align}
which can be written as
\begin{gather}
	\p_\tau^2 \hA^1 - \Delta_s \hA^1 = 0,
	\label{eomhA1}
	\\ \p_\tau^2 \hA^I - \Delta_s \hA^I + 2a e^{2a\xi}\p_I \hA^1 = 0,
	\label{eomhAI}
\end{gather}
where $I=2,3$.
The commutation relation \eqref{transcomm} and the equation of motion \eqref{eomhA1} can be used to write down the
	normal mode expansion of the transverse gauge field $\hA_1$ and its conjugate momentum $\hat \Pi^1$:
\begin{align}
	\hA_1(\tau,\xi,\V x_\perp)
	&= \int\frac{d\w}{\sqrt{2\w}}\frac{d^2k_\perp}{2\pi}\frac{a^2}{\w k_\perp}
		\left[
			a_1(\w,\V k_\perp) e^{-i\w\tau}e^{i\V k_\perp\cdot \V x_\perp} + \hc
		\right]z^2 k_{i\frac{\w}{a}}(z),
		\label{xiA}
	\\ \hat \Pi^1(\tau, \xi, \V x_\perp)
		&= -i\int\frac{d\w}{\sqrt{2\w}}\frac{d^2k_\perp}{2\pi}k_\perp
		\left[
			a_1(\w,\V k_\perp) e^{-i\w\tau}e^{i\V k_\perp\cdot \V x_\perp} - \hc
		\right] k_{i\frac{\w}{a}}(z).
		\label{xiPi}
\end{align}
Requiring that the creation/annihilation operators satisfy the standard commutation relation
\begin{align}
	[a_1(\w,\V k_\perp), a_1^\dagger(\w', \V k'_\perp)] = \delta(\w-\w')\delta^{(2)}(\V k_\perp - \V k'_\perp),
	\label{standardcommrel}
\end{align}
one can readily check that the transverse fields \eqref{xiA}, \eqref{xiPi} satisfy the relation \eqref{transcomm}.
The remaining components $\hA_I$ can also be obtained by writing down a relation similar to \eqref{relation}
	and using the equations of motion \eqref{eomhAI}.
The fields $\hA_I$ has been worked out in \cite{Lenz:2008vw}, the result of which we state here for later reference:
\begin{align}
	\hA_I(\tau,\xi,\V x_\perp)
	&= \int\frac{d\w}{\sqrt{2\w}}\frac{d^2k_\perp}{2\pi}
		\nonumber \\&\quad\times
		\left[
			\left\{
				\epsilon_I(\V k_\perp) a_2(\w,\V k_\perp)
				+ i\frac{ak_I}{\w k_\perp}a_1(\w,\V k_\perp)z\frac{d}{dz}
			\right\}
			k_{i\frac{\w}{a}}(z)e^{-i\w\tau}e^{i\V k_\perp\cdot \V x_\perp} + \hc
		\right]
		\label{AI},
\end{align}
where $k_I$ denotes the $I$-th component of $\V k_\perp$, and the second pair of creation/annihilation operators satisfy the commutation relations
\begin{gather}
	[a_2(\w,\V k_\perp), a_2^\dagger(\w', \V k'_\perp)] = \delta(\w-\w')\delta^{(2)}(\V k_\perp - \V k'_\perp),
	\\
	[a_1(\w,\V k_\perp), a_2^\dagger(\w', \V k'_\perp)] = 0,
\end{gather}
and the polarization vector $\epsilon_I(\V k_\perp)$ is transverse to $\V k_\perp$,
\begin{align}\label{polprop}
	k_I \epsilon^I(\V k_\perp) = 0.
\end{align}

\subsection{Wilson lines, edge modes and surface charges on the horizon}\label{sec:rind1}

In order to investigate the behavior of the transverse fields near the future Rindler horizon,
	we note that the horizon is parametrized by the advanced time $v=\tau + \xi$ as well as the 2-dimensional plane coordinates $\V x_\perp$.
In terms of the advanced coordinates $(v,\xi,\V x_\perp)$, the Rindler metric is
\begin{align}
	ds^2 = e^{2a\xi}(-dv^2 + 2dvd\xi) + d\V x_\perp^2,
\end{align}
and the transverse gauge field \eqref{xiA} is
\begin{align}
	\hA_1(v,\xi,\V x_\perp)
	&= \int\frac{d\w}{\sqrt{2\w}}\frac{d^2k_\perp}{2\pi}\frac{a^2}{\w k_\perp}
		\left[
			a_1(\w,\V k_\perp) e^{-i\w v+i\w\xi}e^{i\V k_\perp\cdot \V x_\perp} + \hc
		\right]z^2 k_{i\frac{\w}{a}}(z).
	\label{vAxi}
\end{align}
Notice the factors $e^{\pm i\w \xi}$ in the integrand.
As we approach the horizon $\xi\to -\infty$, by virtue of the Riemann-Lebesgue lemma only the leading soft modes contribute to the integral.
Similar to the construction of \cite{Kulish:1970ut, Ware:2013zja}, we can explicitly implement this by replacing $e^{\pm i\w\xi}$
	with a scalar function $\phi(\w)$, which satisfies $\phi(0) = 1$ and has support only in a small neighborhood of $\w = 0$.
With this and the asymptotic form
\begin{align}\label{asymK}
	k_{i\frac{\w}{a}}(z) \sim \frac{\w}{a}\sqrt\frac{2}{\pi}K_0(z)\quad\text{as $z\to 0$,}
\end{align}
we obtain
\begin{align}
	\hA_1(v,\xi,\V x_\perp)
	&\sim \int\frac{d\w}{\sqrt{\pi\w}}\frac{d^2k_\perp}{2\pi}\frac{a}{k_\perp}\phi(\w)
		\left[
			a_1(\w,\V k_\perp) e^{i\V k_\perp\cdot \V x_\perp} + \hc
		\right]z^2 K_0(z)
		\quad \text{as $\xi\to -\infty$.}
	\label{asymA}
\end{align}
Now, let us consider the exponent in the Wilson line, i.e., the line integral
\begin{align}
	\mA(x) = \int_\Gamma dx^\mu \hA_\mu(x),
\end{align}
where $\Gamma$ is a time-like path in the vicinity of the horizon. In the following we will evaluate this assuming that the gauge fields satisfy sourceless, quasi-free equations of motion, i.e., we will not consider its interactions with any currents, classical or otherwise. As we have seen in section \ref{sec:mink1}, the current for the straight line asymptotic path is essentially classical. Use of the Yang-Feldman equation \eqref{Yang} in the evaluation of the line integral implies that the interaction terms with the external current gives additional c-number terms in the expression for the Wilson line, which are related to the Coulomb phase. These were not relevant for the analysis of the soft hair for that case and we think it is safe to assume this to be the case here as well.
From the metric \eqref{metric}, one can see that $d\V x_\perp^2 = 0$ along a time-like geodesic as $\xi\to -\infty$.
Thus we may write
\begin{align}
	\mA(\V x_\perp) &= \int_\Gamma d\xi \hA_1(v,\xi,\V x_\perp)
	\\ &=
		-\int\frac{d\w}{\sqrt{\pi\w}}\frac{d^2k_\perp}{2\pi}\frac{1}{k_\perp}\phi(\w)
		\left[
			a_1(\w,\V k_\perp) e^{i\V k_\perp\cdot \V x_\perp} + \hc
		\right],
		\label{mA}
\end{align}
where we used a boundary condition analogous to \eqref{boundarycondition},
\begin{align}\label{Kbdry}
	\int_\Gamma dz\,z K_0(z) = -z K_1(z),
\end{align}
and took the limit corresponding to $\xi\to -\infty$,
\begin{align}\label{limit}
	\lim_{z\to 0}z K_1(z) = 1.
\end{align}
Drawing analogy from Minkowski spacetime, we expect the Wilson line $\exp\left\{ie\mA(x)\right\}$ to serve as the
	Faddeev-Kulish dressings in Rindler spacetime.
To see this, consider any function $\vep(\V x_\perp)$ on the 2-dimensional plane.
The conserved charge $Q_\vep$ associated with this function is then \cite{Strominger:2017zoo}
\begin{align}
	Q_\vep = Q^\text{soft}_\vep + Q^\text{hard}_\vep,
\end{align}
where the soft and hard charges are given as
\begin{align}
	Q^\text{soft}_\vep = \int_H \rmd\vep\wedge *F,\qquad Q^\text{hard}_\vep = \int_H \vep *j.
\end{align}
Here $H$ denotes the future Rindler horizon, $j$ is the charged matter current, and $*F$ is the dual field strength tensor
\begin{align}
	(*F)_\mn = \frac{1}{2}\epsilon_\mnrs F^\rs,
\end{align}
where $\epsilon_\mnrs$ is the antisymmetric Levi-Civita tensor with $\epsilon_{0123} = \sqrt{-g} = e^{2a\xi}$.
Since the operator $\mA(x)$ involves only the soft photon modes (cf. \eqref{mA}),
	it commutes with the hard charge $Q_\vep^\text{hard}$ and we may thus focus our attention on the soft charge $Q^\text{soft}_\vep$.
The horizon is parametrized by $(v,\V x_\perp)$ and $\p_v\vep(\V x_\perp) = 0$,
	which implies that the relevant components of the dual tensor $*F$ are, up to some magnetic fields that vanish at the future horizon,
\begin{align}
	(*F)_{02} = -\p_v\hA^3,
	\qquad (*F)_{03} = \p_v\hA^2.
\end{align}
We therefore have
\begin{align}
	Q^\text{soft}_\vep
	&= -\lim_{\xi\to -\infty}\int_{-\infty}^\infty dv\int d^2\V x_\perp 
		\p_I\vep(\V x_\perp) \p_v \hA^I (v,\xi,\V x_\perp)
	\\&= \int d^2\V x_\perp \vep(\V x_\perp)
		N(\V x_\perp),
\end{align}
where, in the last line we defined
\begin{align}
	N(\V x_\perp) = \lim_{\xi\to -\infty}\int_{-\infty}^\infty dv\,\p_v \p_I\hA^I(v,\xi,\V x_\perp),
\end{align}
after an integration by parts.
The transverse property \eqref{properties} of the projection operator implies that $\p_i \hA^i = 0$,
	or equivalently $\p_I\hA^I = -\p_\xi\hA^1 = -az\p_z\hA^1$.
Thus with the integral representation \eqref{intrepdelta} of the Dirac delta function,
	we obtain
\begin{align}
	N(\V x_\perp)
	&= \lim_{z \to 0}
		i\int\frac{d\w}{\sqrt{2\w}}d^2k_\perp k_\perp\delta(\w)
		\left[
			a_1(\w,\V k_\perp) e^{i\V k_\perp\cdot \V x_\perp} - \hc
		\right] az\frac{d}{dz} k_{i\frac{\w}{a}}(z).
	\\ &= -i\int\frac{d\w}{\sqrt{\pi\w}}d^2k_\perp k_\perp\w\delta(\w)
		\left[
			a_1(\w,\V k_\perp) e^{i\V k_\perp\cdot \V x_\perp} - \hc
		\right],
\end{align}
where in the second line we used the asymptotic form \eqref{asymK}, along with the relation
\begin{align}\label{derivative}
	\frac{d}{dz}K_0(z) = -K_1(z),
\end{align}
and then took the limit $z\to 0$ using \eqref{limit}.
With the commutation relation \eqref{standardcommrel}, we obtain
\begin{align}
	\left[N(\V x_\perp), \mA(\V x_\perp')\right]
	&= i\int\frac{d\w d\w'}{\pi\sqrt{\w\w'}}\frac{d^2k_\perp d^2k'_\perp}{2\pi}\frac{k_\perp}{k'_\perp}
		\w\delta(\w)\phi(\w')
		\nonumber \\ &\qquad\times
		\left[
			a_1(\w,\V k_\perp) e^{i\V k_\perp\cdot \V x_\perp}
			- \hc,
			\ 
			a_1(\w',\V k'_\perp) e^{i\V k'_\perp\cdot \V x'_\perp}
			+ \hc
		\right]
	\\ &= i\int\frac{d\w d\w'}{\pi\sqrt{\w\w'}}\frac{d^2k_\perp d^2k'_\perp}{2\pi}
		\w\delta(\w)\phi(\w')
		\nonumber \\ &\qquad\times
		\delta(\w')\delta^{(2)}(\V k_\perp - \V k'_\perp)
		\left[
			e^{i\V k_\perp \cdot (\V x_\perp - \V x'_\perp)}
			+ e^{-i\V k_\perp \cdot (\V x_\perp - \V x'_\perp)}
		\right]
	\\ &= 2i\delta^{(2)}(\V x_\perp - \V x'_\perp)\int_0^\infty d\w\,\delta(\w)
	\\ &= i\delta^{(2)}(\V x_\perp - \V x'_\perp),
		\label{conj}
\end{align}
where we used the following convention of delta functions,
\begin{align}\label{delta}
	\int_0^\infty d\w\,\delta(\w)f(\w) = \frac{1}{2}f(0).
\end{align}
Equation \eqref{conj} shows that $N(\V x_\perp)$ and $\mA(\V x_\perp)$ are canonically conjugate variables,
	and therefore $\mA(\V x_\perp)$ satisfies the commutator
\begin{align}\label{QmA}
	\left[
		Q_\vep,
		\mA(\V x_\perp)
	\right]
	= i \vep(\V x_\perp).
\end{align}
This has immediate consequence.
Consider the following state,
\begin{align}
	\ket{q,\V x_\perp} = e^{iq\mA(\V x_\perp)}\ket{0},
\end{align}
where we choose $\ket{0}$ to be the vacuum satisfying $Q_\vep\ket{0} = 0$.
Since $\mA(\V x_\perp)$ only involves zero-energy photon operators, this state carries zero energy and is therefore a vacuum.
However, it follows from \eqref{QmA} that
\begin{align}
	Q_\vep \ket{q, \V x_\perp}
	= Q^\text{soft}_\vep \ket{q, \V x_\perp}
	= -q\vep(\V x_\perp) \ket{q,\V x_\perp}.
\end{align}
which implies that this is a degenerate vacuum carrying soft charge.
This is analogous to the case of flat space \cite{Kapec:2017tkm,Choi:2017ylo}, where the set of degenerate vacua is obtained
	by dressing the vacuum with the Faddeev-Kulish operators.
We are thus led to the conclusion that the time-like Wilson lines near the horizon are the Faddeev-Kulish dressings of Rindler spacetime.

We end this section with an instructive derivation of the boundary values of gauge fields
	to see how the charge $Q_\vep$ acts on them.
Let us define the boundary fields
\begin{align}
	\hA^H_i
	&= \lim_{\xi\to-\infty} \hA_i(v,\xi,\V x_\perp).
\end{align}
From \eqref{asymA} we can see that $\hA^H_1=0$ since
\begin{align}
	\lim_{z\to 0}z^2K_0(z) =0.
\end{align}
In the advanced coordinates the remaining components \eqref{AI} can be written as
\begin{align}
	\hA_I(v,\xi,\V x_\perp)
	&= \int\frac{d\w}{\sqrt{2\w}}\frac{d^2k_\perp}{2\pi}
		\left[
			\left\{
				\epsilon_I a_2
				+ i\frac{ak_I}{\w k_\perp}a_1 z\frac{d}{dz}
			\right\}
			k_{i\frac{\w}{a}}(z)e^{-i\w v+i\w\xi}e^{i\V k_\perp\cdot \V x_\perp} + \hc
		\right].
\end{align}
The first term in the curly brackets involving the polarization tensor $\epsilon_I$ is proportional to the expression
\begin{align}
	e^{i\w\xi}k_{i\frac{\w}{a}}(z) = \frac{\w}{a}\sqrt{\frac{2}{\pi}}K_0(z) + O(\w^2) \to 0
	\quad\text{as $\w\to 0$},
	\label{kvanish}
\end{align}
so it retains no zero-energy modes at the horizon.
However, the second term survives, yielding the non-zero components
\begin{align}
	\hA^H_I(\V x_\perp)
	&= -i\int\frac{d\w}{\sqrt{\pi\w}}\frac{d^2k_\perp}{2\pi}\frac{k_I}{k_\perp}\phi(\w)
		\left[
			a_1(\w,\V k_\perp) e^{i\V k_\perp\cdot \V x_\perp} - \hc
		\right].
\end{align}
The expression \eqref{mA} tells us that we can write these fields in terms of $\mA(\V x_\perp)$ as
\begin{align}
	\hA^H_I(\V x_\perp) = \p_I \mA(\V x_\perp),
\end{align}
from which we obtain the commutation relation
\begin{align}
	\left[
		Q_\vep,
		\hA^H_I(\V x_\perp)
	\right]
	= i\p_I \vep(\V x_\perp).
\end{align}
This is reminiscent of the action of charge on boundary fields in Minkowski spacetime \cite{He:2014cra},
\begin{align}
	\left[Q_\vep, A_z(u,z,\zb)\right] = i\p_z \vep(z,\zb),
\end{align}
and shows that $Q_\vep$ correctly generates the boundary degrees of freedom (large gauge transformations), which in our case are the fields at the Rindler horizon.

In summary, we have shown that, similar to the Minkowski spacetime, the Wilson line puncture on the future Rindler horizon carries
	a definite soft horizon charge.
This identifies the puncture as the Faddeev-Kulish dressing of Rindler spacetime, which can be used to generate the edge Hilbert space.
The edge Hilbert space consists of an infinite number of degenerate vacua, where each state is labeled by its soft horizon charge.
This result is consistent with the analysis made in \cite{Blommaert:2018rsf} with regards to the edge states in the Lorenz gauge.
In the next section, we apply similar methods to extend our analysis to the Schwarzschild spacetime and its horizon.

\section{Soft hair on Schwarzschild horizon}\label{sec:schwarzschild}

In this section, we investigate the soft photon hair directly on the Schwarzschild horizon using the quantization method of \cite{Lenz:2008vw}.
The Schwarzschild metric reads
\begin{align}\label{fullSchwarzschild}
	ds^2 = -\left(1-\frac{2M}{r}\right)dt^2 + \left(1-\frac{2M}{r}\right)^{-1}dr^2 + r^2 \left(d\theta^2 + \sin^2\theta d\phi^2\right),
\end{align}
where $M= GM_0$, with $G$ the Newton's constant and $M_0$ the mass of the black hole.
From the lessons learned in Rindler spacetime, we know that it is only the near-horizon physics that plays a role in the analysis:
	it is expected that the Wilson lines along a time-like path in the vicinity of the horizon will again be the analogue of the Faddeev-Kulish dressings,
	the building blocks of the edge Hilbert space.
This motivates us to restrict our attention to the near-horizon region of Schwarzschild, by writing
\begin{align}
	\rho = r - 2M.
\end{align}
The leading terms in the small-$\rho$ expansion of \eqref{fullSchwarzschild} yields the near-horizon metric,
\begin{align}
	ds^2 = -\frac{\rho}{2M}dt^2 + \frac{2M}{\rho}d\rho^2 + 4M^2\left(d\theta^2 + \sin^2\theta d\phi^2\right).
\end{align}
Then, let us define new coordinates $\xi$ and $y$ as
\begin{align}
	\xi = 2M\ln\left(\frac{\rho}{2M}\right),\qquad y = \cos\theta,
\end{align}
in terms of which the metric reads
\begin{align}\label{nhm}
	ds^2 = e^{2a\xi}\left(-dt^2 + d\xi^2\right) + 4M^2\left[\frac{dy^2}{1-y^2} + (1-y^2)d\phi^2\right],
\end{align}
where $a=\frac{1}{4M}$ is the surface gravity of the black hole.
We will use $\Omega$ to denote the spherical coordinates $(y,\phi)$ collectively.
The $(t,\xi)$ space resembles the Rindler spacetime, but the remaining space is a 2-sphere, not a 2-plane.
There are two reasons for choosing the coordinate $y=\cos\theta$ over the conventional $\theta$:
One is that this achieves $\p_i(\sqrt{-g}g^{00}) = 0$ which simplifies a lot of calculations \cite{Lenz:2008vw},
	and the other is that it is difficult to obtain a simple operator relation such as \eqref{relation}
	if we use $\theta$.

In the following subsections, we will be working with the metric \eqref{nhm}.
We will begin by deriving the mode expansion of transverse gauge fields.
These will be used to show that the near-horizon, time-like Wilson lines are the Faddeev-Kulish dressings
	that build the edge Hilbert space.

\subsection{Transverse gauge fields}

We will work in the Weyl gauge as before,
\begin{align}
	A_0(t,\xi,\Omega) = 0.
\end{align}
With the metric \eqref{nhm}, the transverse projection operator is defined as,
\begin{align}
	{P^i}_j = \delta^i_j - \p^i\frac{1}{\Delta}\p_j,
\end{align}
where the operator $\Delta = \p_i\p^i$ is now given by
\begin{align}
	\Delta
	&= \p_\xi e^{-2a\xi} \p_\xi - \frac{\V L^2}{4M^2}.
\end{align}
Here $\V L^2$ is the angular momentum squared operator,
\begin{align}
	\V L^2 = -\p_y (1-y^2)\p_y - \frac{\p_\phi^2}{1-y^2},
\end{align}
whose eigenfunctions are the spherical harmonics $Y_{\ell m}$,
\begin{align}
	\V L^2 Y_{\ell m}(y,\phi) = \ell(\ell + 1) Y_{\ell m}(y,\phi).
\end{align}
In order to find the mode expansion of transverse gauge fields, as for the Rindler case, let us first look at the case of free scalar field $\varphi$,
	whose equation of motion has the form
\begin{align}
	\frac{1}{\sqrt{-g}}\p_\mu \sqrt{-g}g^\mn \p_\nu \varphi = 0,
\end{align}
or equivalently,
\begin{align}
	\label{schwarzschildeom}
	\left(\p_t^2 - \Delta_s\right)\varphi = 0,
\end{align}
where $\Delta_s$ is the scalar Laplacian,
\begin{align}
	\Delta_s &= \p_\xi^2 - \frac{e^{2a \xi}}{4M^2}\V L^2.
\end{align}
From symmetry, solutions take the form
\begin{align}
	\varphi(t,\xi,y,\phi) = e^{-i\w t} Y_{\ell m}(y,\phi) R(\xi)
\end{align}
with some function $R$ of $\xi$.
Then, the equation of motion \eqref{schwarzschildeom} reduces to an equation for $R$,
	which reads
\begin{align}\label{eqnR}
	\frac{d^2R}{d\xi^2} + \left[\w^2 - \frac{e^{2a\xi}}{4M^2}\ell(\ell+1)\right]R = 0.
\end{align}
Now, define a new variable
\begin{align}
	z = 2\sqrt{\ell(\ell+1)}e^{a\xi},
\end{align}
where we exclude the $\ell=0$ mode. We will see later that $\ell=0$ is associated to the total electric charge
	and hence is not of our interest.
In terms of $z$, equation \eqref{eqnR} becomes
\begin{align}
	z^2\frac{d^2R}{dz^2} + z\frac{dR}{dz} + \left(\frac{\w^2}{a^2} - z^2\right)R = 0.
\end{align}
This is the same modified Bessel equation that we saw in Rindler space, whose solutions are the properly normalized MacDonald functions
\begin{align}
	R = k_{i\frac{\w}{a}}(z) = \frac{1}{\pi}\sqrt{\frac{2\w}{a}\sinh\left(\frac{\pi \w}{a}\right)}K_{i\frac{\w}{a}}(z).
\end{align}

Now we're in the right place to obtain the commutation relation of the transverse fields.
From the free Maxwell Lagrangian density,
\begin{align}
	\mL = \sqrt{-g}\left(-\frac{1}{4}F^\mn F_\mn\right),
\end{align}
we obtain the momentum density $\Pi^i$ conjugate to the gauge field to be
\begin{align}
	\Pi^i = -\sqrt{-g}g^{00}\p_0 A^i = 4M^2 \p_0 A^i.
\end{align}
Then we quantize the fields by imposing the equal-time commutation relation,
\begin{align}\label{schwarzschildetcr}
	\left[\Pi^i(t,\xi,y,\phi), A_j(t,\xi',y',\phi')\right]
	&= \frac{1}{i}\delta^i_j\delta(\xi-\xi')\delta(y-y')\delta(\phi-\phi').
\end{align}
Commutation relation between transverse components of the fields is obtained by applying transverse projection onto \eqref{schwarzschildetcr}.
The projection operator relevant to our analysis is ${P^1}_1$.
Observing that
\begin{align}
	\Delta_s \left(1 - \p^1\Di \p_1\right) = -\frac{e^{2a\xi}}{4M^2}\V L^2,
\end{align}
we find that the projection operator can be written as
\begin{align}
	{P^1}_1 = -\Dis \frac{e^{2a\xi}}{4M^2}\V L^2.
\end{align}
Then, we obtain
\begin{align}
	\left[
		\hat \Pi^1(t,\V x'),
		\hA_1(t,\V x)
	\right]
	&= \frac{1}{i}g_{11}{P^1}_1 g^{11}\delta(\xi-\xi')\delta(y-y')\delta(\phi-\phi')
	\\ &= i\frac{e^{2a\xi}}{4M^2}\Dis \V L^2 \delta(\xi-\xi')\delta(y-y')\delta(\phi-\phi'),
\end{align}
which, making use of the completeness relations
\begin{gather}
	\int_0^\infty d\w\, k_{i\frac{\w}{a}}(z) k_{i\frac{\w}{a}}(z') = \delta(\xi - \xi'),
	\\
	\sum_{\ell=0}^\infty \sum_{m=-\ell}^\ell Y_{\ell m}(y,\phi) Y^*_{\ell m}(y', \phi') = \delta(y-y')\delta(\phi-\phi'),
	\label{completenesssphericalharmonics}
\end{gather}
can be written as
\begin{align}
	\left[
		\hat \Pi^1(t,\V x'),
		\hA_1(t,\V x)
	\right]
	= i\sum_{\ell m}z^2 Y_{\ell m}(y,\phi) Y^*_{\ell m}(y',\phi')\int_0^\infty d\w \frac{a^2}{\w^2} k_{i\frac{\w}{a}}(z) k_{i\frac{\w}{a}}(z').
	\label{etcrschwarzschild1}
\end{align}
Now that we have the commutation relation, let us consider the equation of motion of free gauge field in Weyl gauge, which reads
\begin{align}\label{4eomschwarzschild}
	\p_0 \sqrt{-g}g^{00}\p_0 A^i
		+ \p_j \sqrt{-g}g^{jk}g^{il}(\p_k A_l - \p_l A_k) = 0.
\end{align}
Noting that $\p_i A_j - \p_jA_i = \p_i\hA_j - \p_j\hA_i$, one may write the equations of motion of the transverse fields as
\begin{align}
	\sqrt{-g}g^{00}\p^2_0 \hA^i
		+ \p_j \sqrt{-g}g^{jk}g^{il}(\p_k \hA_l - \p_l \hA_k) = 0,
\end{align}
or written out explicitly,
\begin{gather}
	-\p_0^2 \hA^1 + \Delta_s\hA^1 = 0,
	\label{eomschwarzschild1}
	\\
	- \p_0^2 \hA^y + \Delta_s\hA^y
		+ \frac{e^{2a\xi}}{4M^2}\left[
			2y\p_\phi \hA^\phi
			-2a\left(1-y^2\right)\p_y \hA^1
			+ 2y\p_y \hA^y
		\right]=0,
	\label{eomschwarzschild2}
	\\
	- \p_0^2 \hA^\phi + \Delta_s\hA^\phi
		- \frac{e^{2a\xi}}{4M^2}\left[
			\p_y \left(2y\hA^\phi\right)
			+\frac{2a\p_\phi \hA^1}{1-y^2}
			+ \frac{2y\p_\phi \hA^y}{(1-y^2)^2}
		\right]=0.
	\label{eomschwarzschild3}
\end{gather}
Using the equal-time commutation relation \eqref{etcrschwarzschild1} and the equation of motion \eqref{eomschwarzschild1},
	we obtain the $\xi$-component of the transverse fields to be
\begin{align}
	\hA^1(t,\xi,\Omega)
	&= \sum_{\ell m}4\sqrt{\ell(\ell+1)}\int\frac{d\w}{\sqrt{2\w}}\frac{a^2}{\w}
		\left[
			a_{\ell m}(\w) e^{-i\w t}Y_{\ell m}(\Omega)
			+ \hc
		\right]k_{i\frac{\w}{a}}(z),
	\label{trans1}
	\\
	\hA_1(t,\xi,\Omega) &= \sum_{\ell m}\frac{1}{\sqrt{\ell(\ell+1)}}\int\frac{d\w}{\sqrt{2\w}}\frac{a^2}{\w}
		\left[
			a_{\ell m}(\w) e^{-i\w t}Y_{\ell m}(\Omega)
			+ \hc
		\right]z^2k_{i\frac{\w}{a}}(z).
	\label{trans2}
\end{align}
Due to the property $\p_i(\sqrt {-g}g^{00})=0$ of our metric \eqref{nhm}, the transverse conjugate momentum is simply
\begin{align}
	\hat \Pi^1(t,\xi,\Omega)
	&=4M^2\p_0\hA^1(t,\xi,\Omega)
	\\ &= -i\sum_{\ell m}\sqrt{\ell(\ell+1)}\int\frac{d\w}{\sqrt{2\w}}
		\left[
			a_{\ell m}(\w) e^{-i\w t}Y_{\ell m}(\Omega)
			- \hc
		\right]k_{i\frac{\w}{a}}(z).
	\label{trans3}
\end{align}
One can readily check that by postulating the standard commutation relation
\begin{align}
	\left[
		a_{\ell m}(\w),
		a_{\ell' m'}^\dagger(\w')
	\right]
	= \delta_{\ell\ell'}\delta_{mm'}\delta(\w-\w'),
\end{align}
the transverse fields \eqref{trans2} and \eqref{trans3} satisfy \eqref{etcrschwarzschild1}.

\subsection{Wilson lines and edge degrees of freedom}

Since the horizon is parametrized by the 2-sphere coordinates $(y,\phi)$ and the advanced time $v=t+\xi$,
	let us change coordinates to $(v,\xi,y,\phi)$.
In terms of these coordinates, the radial field \eqref{trans2} is simply obtained by replacing $t$ with $v-\xi$,
\begin{align}
	\hA_1(v,\xi,\Omega)
	&= \sum_{\ell m}\frac{1}{\sqrt{\ell(\ell+1)}}\int\frac{d\w}{\sqrt{2\w}}\frac{a^2}{\w}
		\left[
			a_{\ell m}(\w) e^{-i\w v + i\w\xi}Y_{\ell m}(\Omega)
			+ \hc
		\right]z^2k_{i\frac{\w}{a}}(z).
\end{align}
In the vicinity of the horizon $\xi\to -\infty$, only the leading soft modes contribute to the integral,
	due to the factor $e^{\pm i\w\xi}$ and the Riemann-Lebesgue lemma.
As in \cite{Kulish:1970ut,Ware:2013zja}, we implement this with a scalar function $\phi(\w)$ that has support in a small neighborhood
	of $\w=0$ and satisfies $\phi(0)=1$:
\begin{align}
	\hA_1(v,\xi,\Omega)
	&= \sum_{\ell m}\frac{1}{\sqrt{\ell(\ell+1)}}\int\frac{d\w}{\sqrt{2\w}}\frac{a^2}{\w}
		\phi(\w)\left[
			a_{\ell m}(\w) Y_{\ell m}(\Omega)
			+ \hc
		\right]z^2k_{i\frac{\w}{a}}(z).
\end{align}
Let us consider the line integral
\begin{align}
	\mA(x) = \int^x_\Gamma dz^\mu \hA_\mu(z),
\end{align}
where $\Gamma$ is a time-like path in the vicinity of the horizon. We are again treating the gauge fields to satisfy sourceless quasi-free equations of motion, for the same reason discussed in section \ref{sec:rind1}.
From the metric \eqref{nhm}, one can observe that $dy,d\phi\to 0$ along $\Gamma$ as $\xi\to -\infty$,
	which allows us to write
\begin{align}
	\mA(\Omega)
	&= \int_\Gamma d\xi \hA_1(v,\xi,\Omega)
	\\ &=-\sum_{\ell m}\frac{1}{\sqrt{\ell(\ell+1)}}\int\frac{d\w}{\sqrt{\pi\w}}\phi(\w)
		\left[
			a_{\ell m}(\w)Y_{\ell m}(\Omega)
			+ \hc
		\right]
		\label{Aschwarzschild}
\end{align}
where we used the asymptotic form \eqref{asymK} of the MacDonald function and the boundary condition \eqref{Kbdry}.

As in the case of Rindler spacetime, we want to show that the Wilson line $\exp\left\{ie\mA(x)\right\}$ is the
	Faddeev-Kulish dressing that implements soft hair on the Schwarzschild horizon.
To this end, let us consider a 2-sphere function $\vep(\Omega)$.
The conserved charge $Q_\vep$ of QED associated with $\vep(\Omega)$ is \cite{Strominger:2017zoo}
\begin{align}
	Q_\vep = Q^\text{soft}_\vep + Q^\text{hard}_\vep,
\end{align}
where we have
\begin{align}
	Q^\text{soft}_\vep = \int_H \rmd\vep\wedge *F,\qquad Q^\text{hard}_\vep = \int_H \vep *j,
\end{align}
with the Schwarzschild horizon $H$ and the charged matter current $j$.
Equation \eqref{Aschwarzschild} shows that $\mA(\Omega)$ only involves zero-energy photon operators, which implies
	that it commutes with the hard charge $Q^\text{hard}_\vep$.
To obtain an explicit expression for the soft charge $Q^\text{soft}_\vep$, we note that the horizon $H$ is parametrized by $v$, $y$, and $\phi$.
Thus the two relevant components of the dual field tensor are, up to some magnetic fields that vanish at $H$,
\begin{align}
	(*F)_{vy} = -4M^2\p_v\hA^\phi,\qquad
	(*F)_{v\phi} = 4M^2\p_v\hA^y,
\end{align}
using which we may write the soft charge $Q_\vep^\text{soft}$ as
\begin{align}
	Q^\text{soft}_\vep &= -4M^2\int_{-\infty}^\infty dv\int_{-1}^1 dy\int_0^{2\pi} d\phi\,
		\Big \{
			\p_y\vep(y, \phi) \p_v \hA^y
			+ \p_\phi\vep(y, \phi) \p_v\hA^\phi
		\Big\}.
\end{align}
After a partial integration, we may write
\begin{align}
	Q^\text{soft}_\vep &= \int d\Omega \,\vep(\Omega) N(\Omega),
	\label{softchargeN}
\end{align}
where we defined the operator $N(\Omega)$ as
\begin{align}
	N(\Omega) &= 4M^2\int_{-\infty}^\infty dv\,\p_v (\p_y\hA^y + \p_\phi\hA^\phi).
\end{align}
Using the property $\p_i \hA^i=0$ of transverse fields and the definition $a=1/4M$, we obtain
\begin{align}
	N(\Omega)
	&= -\frac{1}{4a^2}\int_{-\infty}^\infty dv\,\p_v \p_1 \hA^1.
\end{align}
Now, we can substitute the mode expansion \eqref{trans1} and use \eqref{asymK}, \eqref{derivative}
	as well as the integral representation \eqref{intrepdelta} to write $N(\Omega)$ in terms of the zero-mode photon operators,
\begin{align}
	N(\Omega)
	&= -4i \sum_{\ell m}\sqrt{\ell(\ell+1)}\int\frac{d\w}{\sqrt{\pi\w}}
		\w\delta(\w)
		\left[
			a_{\ell m}(\w) Y_{\ell m}(\Omega)
			- \hc
		\right],
	\label{Nschwarzschild}
\end{align}
where we also took the limit \eqref{limit} since the gauge field is evaluated at the horizon.
Now that we have expressions \eqref{Aschwarzschild} and \eqref{Nschwarzschild}, one can see by direct calculation
	that $N(\Omega)$ and $\mA(\Omega)$ are conjugate variables, up to a constant,
\begin{align}
	\left[N(\Omega),\mA(\Omega')\right]
	&= 2i \sum_{\ell m}\sum_{\ell'm'}\sqrt{\frac{\ell(\ell+1)}{\ell'(\ell'+1)}}
		\int\frac{d\w d\w'}{\sqrt{\w\w'}} \w\delta(\w)\phi(\w')
		\nonumber \\&\quad\times
		\left[
			a_{\ell m}(\w) Y_{\ell m}(\Omega)
			- \hc\,,\,
			a_{\ell'm'}(\w')Y_{\ell'm'}(\Omega')
			+ \hc
		\right]
	\\ &= i \sum_{\ell=1}^\infty\sum_{m=-\ell}^\ell
		Y_{\ell m}(\Omega)Y_{\ell m}^*(\Omega')
	\\ &= i\delta(y - y')\delta(\phi - \phi') - \frac{i}{4\pi},
	\label{cr_with_const}
\end{align}
where we used the convention \eqref{delta} of delta function and the completeness relation of spherical harmonics
	\eqref{completenesssphericalharmonics}.
If we expand the gauge parameter $\vep(\Omega)$ in spherical harmonics, the $\ell=0$ mode is associated
	to the conservation of total electric charge \cite{Strominger:2017zoo}, which is not of our interest.
Thus we want to restrict our attention to the case where
\begin{align}
	\int d\Omega\, \vep(\Omega) = 0.
\end{align}
Then it follows from \eqref{softchargeN} that the following commutation relation is satisfied,
\begin{align}
	\label{finalcr}
	\left[Q_\vep,\mA(\Omega)\right] = i\vep(\Omega).
\end{align}
The implication of this is that the Wilson lines $e^{ie\mA}$
	are indeed the Faddeev-Kulish dressings corresponding to the soft hair residing at the Schwarzschild horizon.
To illustrate this point, let $\ket{M}$ denote a state describing a Schwarzschild black hole with no soft hair.
Since Schwarzschild black holes carry no electromagnetic charge, $Q_\vep\ket{M}=0$.
Let us construct another state by dressing $\ket{M}$ with a Wilson line,
\begin{align}
	\ket{M,(q,\Omega)} = e^{iq\mA(\Omega)}\ket{M}.
\end{align}
From \eqref{Aschwarzschild} we can see that the operator $\mA(\Omega)$ only involves soft photons;
	the dressing carries no additional energy, angular momentum, or electromagnetic charge.
Unlike $\ket{M}$, however, this new state carries soft hair on the horizon,
\begin{align}
	Q_\vep \ket{M,(q,\Omega)} = \left[Q_\vep, e^{iq\mA(\Omega)}\right]\ket{M} = -q\vep(\Omega)\ket{M,(q,\Omega)}.
\end{align}
This implies that there exists an infinite number of such degenerate states, each labeled by its soft charge configuration.
Given a quantum black hole state with soft hair, one can shift its soft charge using a Wilson line operator.
	
We end the section by analyzing the action of $Q_\vep$ on the boundary gauge fields, given by
\begin{align}
	&\hA^H_y(y,\phi) \equiv \lim_{\xi\to -\infty}\hA_y(v,\xi,y,\phi),\\
	&\hA^H_\phi(y,\phi) \equiv \lim_{\xi\to -\infty}\hA_\phi(v,\xi,y,\phi).
\end{align}
Since these fields are purely large-gauge,
	they can be obtained indirectly via the relations $\hA^H_y(y,\phi) = \p_y\mA(y,\phi)$ and $\hA^H_\phi(y,\phi) = \p_\phi\mA(y,\phi)$.\footnote{
	This can be explicitly shown by obtaining a mode expansion of the transverse fields using the equations of motion
	\eqref{eomschwarzschild2} and \eqref{eomschwarzschild3} and then taking the limit $\xi\to -\infty$.
	As an example, a derivation for $\hA_y^H$ is done in \cref{appendix_Ay}.}
Under a large gauge transformation $\delta \hA_i = \p_i\vep$, the commutation relation \eqref{finalcr}
	implies
\begin{align}
	&\left[Q_\vep,\hA^H_y(y,\phi)\right] = i\p_y\vep(y,\phi) = i\delta \hA^H_y(y,\phi),
	\\ &\left[Q_\vep,\hA^H_\phi(y,\phi)\right] = i\p_\phi\vep(y,\phi) = i\delta \hA^H_\phi(y,\phi).
\end{align}
Therefore, we conclude that the conserved charge $Q_\vep$ correctly generates the boundary degrees of freedom.

To summarize, we have identified the Wilson line punctures on the Schwarzschild horizon as the Faddeev-Kulish dressings
	that carry definite soft horizon charge.
Similar to the case of Minkowski and Rindler spacetimes,
	these dressings can be used to generate the edge Hilbert space consisting of an infinite number of states,
	each of which is labeled by its soft horizon charge.
In this case, the bulk state is a quantum state labeled solely by the mass of the Schwarzschild black hole.
The existence of the edge Hilbert state implies that this bulk state is degenerate, and thus a new quantum number, e.g. the soft horizon charge,
	should be introduced to correctly identify the state.
This is consistent with the Hawking-Perry-Strominger analysis, which claims that the Schwarzschild black holes carry soft hairs
	\cite{Hawking:2016msc,Hawking:2016sgy}.

\section{Discussion}\label{sec:discussion}

In this paper, we have applied the Weyl-gauge quantization scheme of transverse photon fields developed in \cite{Lenz:2008vw}
	to show that for the QED in Rindler and Schwarzschild backgrounds,
	the Wilson line punctures on the horizon are objects that correspond to the Faddeev-Kulish dressings.
By computing the commutation relation between Wilson lines and the soft charge,
	we have shown that each dressing carries a definite soft horizon charge.
The dressings can be used as building blocks to generate the so-called edge Hilbert space
	(as opposed to the bulk Hilbert space),
	that consists of an infinite number of degenerate states each of which is labeled by its charge.
Our work shows that the Wilson line dressing is an effective tool to study the soft hair at both infinity and the horizon.
Moreover, our approach provides for a systematic way to construct the edge Hilbert space which will have applications
	in studies of entanglement entropy of gauge fields \cite{Donnelly:2016auv}.

We have provided a straightforward quantum-mechanical calculation that demonstrates the existence of soft charges localized on
	the Rindler and Schwarzschild horizons,
supports the claim that Schwarzschild black holes carry soft hair \cite{Hawking:2016msc,Hawking:2016sgy},
	and also bridges the gap between the Hawking-Perry-Strominger analysis and the Wilson line formulation \cite{Blommaert:2018oue}
	of Rindler edge states. Moreover, our calculations show explicitly that the limit of gauge fields at the horizon only involve static photons.
This suggests that similar results are expected in curved spacetimes exhibiting an infinite red-shift surface, for example the
	cosmological horizon of a de Sitter space.

A natural step forward would be to extend our analysis to the perturbative quantum gravity in Rindler and Schwarzschild backgrounds,
	via the path-dependent formulation of gravity due to Mandelstam \cite{Mandelstam:1962us,Mandelstam:1968ud}.
	Working directly in a Schwarzschild background, it would be of interest to study the nature of the horizon charges and to investigate the role of supertranslations and superrotations. Another problem of interest is the contribution of higher-spin fields in a Schwarzschild background. Whether massless or massive, these will have soft hair on the horizon. Can one construct an edge Hilbert space in this case in a consistent manner?  Another possible direction would be to study how the Wilson lines and dressings can be analyzed
	in the path-integral formulation of QED and gravity; some recent related references are \cite{Blommaert:2018oue,Wilson-Gerow:2018egh}.
Lastly, while this paper was in preparation we came across a recent paper investigating the effects of gravitational dressings of particles
	at the Schwarzschild horizon \cite{Javadinazhed:2018mle}.
It will be interesting to see how the dressings in \cite{Javadinazhed:2018mle} relate to our construction of Wilson line punctures.

\acknowledgments

We would like to thank Sandeep Pradhan for discussions and Malcolm Perry, Andy Strominger, and Valya Zakharov for useful comments
	on the first draft of the manuscript.
S.C. gratefully acknowledges support from the Leinweber Fellowship and the Samsung Fellowship.

\appendix

\section{Horizon gauge field component}\label{appendix_Ay}

In this appendix we show explicitly that the horizon field defined by
\begin{align}
	\hA^H_y(\Omega) = \lim_{\xi\to-\infty} \hA_y(v,\xi,\Omega),
\end{align}
satisfies the relation $\hA^H_y = \p_y \mA(\Omega)$.
To this end, we want to obtain a mode expansion for $\hA^y(t,\xi,\Omega)$.
Let us take the equation of motion \eqref{eomschwarzschild2} and use $\p_i \hA^i = 0$ to write
\begin{align}
	- \p_0^2 \hA^y + \Delta_s\hA^y
		- \frac{e^{2a\xi}}{4M^2}\left[
			2y\p_1 \hA^1
			-2a\left(1-y^2\right)\p_y \hA^1
		\right]=0.
		\label{eom_appAy}
\end{align}
Now, consider the following ansatz
\begin{align}
	\hA^y &= \hat a^y + (1-y^2)\p_y\frac{1}{\V L^2}\p_1\hA^1,
	\label{ansatzAy}
\end{align}
for some field $\hat a^y$.
Substituting \eqref{ansatzAy} into \eqref{eom_appAy} yields the equation of motion of $\hat a^y$,
	which is essentially that of a free scalar field,
\begin{align}
	\p_0^2 \hat a^y - \Delta_s \hat a^y = 0.
	\label{aeom}
\end{align}
Using equations \eqref{eom_appAy} and \eqref{aeom} along with the mode expansion of $\hA^1$ yields
	the following mode expansion for $\hA^y$,
\begin{align}
	\hA^y(t,\xi,\Omega)
	&= \sum_{\ell m}\int\frac{d\w}{\sqrt{2\w}}
		\Bigg\{
			\Bigg[
				(\cdots)
				+ \frac{4a^3}{\w}\frac{a_{\ell m}(\w)}{\sqrt{\ell(\ell+1)}} (1-y^2)\p_yY_{\ell m}(\Omega)z\frac{d}{dz}k_{i\frac{\w}{a}}(z)
			\Bigg]e^{-i\w t}
			+ \hc
		\Bigg\},
\end{align}
where the omitted terms in the parentheses $(\cdots)$ correspond to the mode expansion for the field $\hat a^y$.
The covariant component $\hA_y$ in terms of the advanced time coordinates $(v,\xi,\Omega)$ is therefore
\begin{align}
	\hA_y(v,\xi,\Omega)
	&= \sum_{\ell m}\int\frac{d\w}{\sqrt{2\w}}
		\Bigg\{
			\Bigg[
				(\cdots)
				+ \frac{a}{\w}\frac{a_{\ell m}(\w)}{\sqrt{\ell(\ell+1)}} \p_yY_{\ell m}(\Omega)z\frac{d}{dz}k_{i\frac{\w}{a}}(z)
			\Bigg]e^{-i\w v + i\w\xi}
			+ \hc
		\Bigg\}.
\end{align}
We can obtain the horizon gauge field $\hA^H_y$ by taking the limit $\xi\to-\infty$.
Since $\hat a^y$ satisfies the free scalar field equation \eqref{aeom}, the terms in $(\cdots)$ are proportional
	to $k_{i\frac{\w}{a}}(z)=O(\w)$ and therefore vanish at the horizon due to the relation \eqref{kvanish}.
Hence,
\begin{align}
	\hA^H_y(\Omega)
	&= \lim_{\xi\to -\infty}\hA_y(v,\xi,\Omega)
	\\ &= -\sum_{\ell m}\int\frac{d\w}{\sqrt{\pi\w}}\phi(\w)
		\Bigg\{
			\frac{a_{\ell m}(\w)}{\sqrt{\ell(\ell+1)}} \p_yY_{\ell m}(\Omega)
			+ \hc
		\Bigg\}.
\end{align}
From \eqref{Aschwarzschild}, we can immediately obtain
\begin{align}
	\hA^H_y(\Omega) = \p_y \mA(\Omega),
\end{align}
which proves the claim.

\bibliographystyle{jhep} 
\bibliography{references}

\end{document}